%% file: closed_loop_DAA_2023.tex
\documentclass[letterpaper, 10 pt, conference]{ieeeconf}  

\IEEEoverridecommandlockouts                              

\usepackage{graphicx}
\usepackage{courier}
\usepackage{tabularx}

\title{\LARGE \bf
Analyzing the Closed-Loop Performance of Detect-And-Avoid Systems
}

\author{\'Italo Romani de Oliveira, Thiago Matsumoto, Aaron Mayne, Antonio Gracia Berna
\thanks{*The authors are with Boeing Research \& Technology, which supported this work. The corresponding author is 
        {\tt\small italo.romanideoliveira@boeing.com}
}
}

\begin{document}

\maketitle
\thispagestyle{empty}
\pagestyle{empty}

\begin{abstract}
        The Detect-And-Avoid (DAA) algorithms for unmanned air vehicles in civil airspace have industry standards called Minimum Operational Performance Standards (MOPS), which establish clear criteria to check whether they can ensure safe separation for all plausible operational conditions. However, these MOPS ensure performance for the avoidance maneuvers, which are open-loop, but not for the maneuvers that bring the air vehicles back to their intended courses after they are deemed clear of conflict, which close the control loop of the missions. In this paper, we analyze the closed-loop performance of existing DAA algorithms that fulfill certain MOPS', by experimenting large numbers of traffic configurations with 4 aircraft in a delimited airspace. 

        We perform this analysis by obtaining their rates of loss of separation and timeout events, the latter happening when a chain of maneuvers exceeds the maximum supply of energy in a vehicle. In pair with these indicators, we measure the efficiency of the closed-loop logic, expressed as the excess fuel rate, and study the relationship between safety and efficiency in these scenarios. Our results show that the inefficiency caused by DAA algorithms is very significant in dense airspaces and that it is necessary to do considerable research in devising closed-loop mission management that takes into account the necessity of acting on DAA advisories and then having a safe and efficient logic to resume to mission.
        
        Performing the simulations of the closed-loop mission management logic can be highly time consuming, depending on the implementation of the DAA algorithm. Despite there being Neural Network approximations of DAA logic that alleviate the computational load, none of those that we found available can properly handle cases with more than one intruder in the ownship surveillance range. Therefore, in an attempt to overcome the high computational cost of analyzing the closed-loop performance of DAA algorithms, we study the correlation between inefficiency and safety of closed-loop scenarios, and some indicators from the corresponding open-loop scenarios, such as angle of deviation and number of deviations per flight time.  
\end{abstract}

\section{INTRODUCTION}
\input{introduction.tex}

\section{MULTIVEHICLE SCENARIO DEFINITIONS}\label{section:scenario_definitions}
\input{scenario_definition.tex}

\section{RESULTS AND ANALYSIS}
\input{analysis.tex}

\section{CONCLUSIONS}
\input{conclusion.tex}

\addtolength{\textheight}{-12cm}   




\section*{ACKNOWLEDGMENT}
We highly appreciate the support of the FAA TCAS Program Office, for authorizing and helping us to obtain the latest version of the ACAS sXu prototype package. Other invaluable contributions were provided by the members of the ACAS sXu development team, including Joshua Silbermann and Tyler Young, who guided and troubleshot the software installation, and Charles Leeper, who taught us basic and advanced features of the software package.


\bibliographystyle{IEEEtran}
\bibliography{./closed_loop_DAA_2023.bib}

\end{document}

%% file: introduction.tex
The social and economic drives for Advanced Air Mobility (AAM) will soon lead to congested skies overhead cities. Thus, higher automation levels will become a requirement for coordinating this traffic, because of its higher complexity, which will not be effectively managed by humans. The establishment of fixed air routes can reduce complexity, however at the cost of limiting traffic capacity and decreasing flight efficiency. Another alternative is the use of a powerful central system equipped with Artificial Intelligence (AI), which would allow flexible trajectories and higher efficiency. However, such system could contain Single-Points-of-Failure (SPoFs), would be a highly sought target of malicious attacks, and would be subject to periods of unavailability. Furthermore, such system would have to impose certain constraints to the trajectories, in order to be computationally feasible, which occasionally may decrease flight efficiency. 

A simpler solution already exists to keep unmanned aircraft safely separated in the sky, called Detect-And-Avoid (or simpy DAA), which addresses the need of avoiding collisions with other aircraft and, in principle, to any detectable flying vehicle. This capability is so essential for future aviation that is already specified in several industry standards, such as DO-365B \cite{DO_365}, DO-386 \cite{DO_386}, DO-396 \cite{DO_396} and others. However, these standards focus on specifying the avoidance maneuver, which maintains separation, and are not concerned with what happens next, when the aircraft are deemed clear of conflict by the DAA algorithm. In other words, they ensure performance of divergent maneuvers, but do not cover the maneuvers needed to converge to a waypoint or terminal point.   

In a previous paper \cite{IROliveira2021}, we analyzed the performance of the DAIDALUS \cite{DaidalusPage} algorithm, which meets the standard DO-365B \cite{DO_365}, as the representative of a fully distributed traffic management strategy, comparing it to two other strategies, one of them fully centralized, and the other one being a distributed, but coordinated strategy. In the follow-on work of that paper, we wanted to explore other versions of DAA and, with that, confirm or disprove our findings, so this paper reports what has been accomplished towards this goal. The closest alternative to DAIDALUS is ACAS-Xu \cite{Manfredi2016,Owen2019}, which not only meets DO-365B \cite{DO_365}, but has a standard of its own, DO-386 \cite{DO_386}, because it includes the capability of processing raw track data from various sensors and doing its own probabilistic data fusion for surveillance, as opposed to \cite{DO_365}, which assumes a unified track/vector information. A systematic comparison between DAIDALUS and ACAS-Xu was already performed in \cite{Davies18}, which pointed to an overall agreement between both, with an expected superiority of ACAS Xu in dealing with noise in the input sensors. However, all the analyses in that reference are concerned only with the open-loop maneuvers, with just one intruder aircraft, and do not consider the resume of the flights towards a destination.

In the present paper, we advance some steps towards generalization of the performance analysis of DAA algorithms in a closed-loop context. We did not have an authorized ACAS Xu software package immediately available to use, but we did obtain authorization from the FAA TCAS Program Office to use the ACAS sXu API (herein version 4.1), which is a variant of ACAS Xu for small Unmanned Air Vehicles (UAVs), limited to 55 pounds of take-off weight, as specified in the standard DO-396 \cite{DO_396}. ACAS sXu inherits the same core algorithm of ACAS Xu, albeit with different look up tables, thus any comparison with DAIDALUS has to take into account their different scopes. Among other differences, the distance-based separation requirement that the aircraft controlled by ACAS sXu has to comply towards manned aircraft or large UAVs is, horizontally, 2,000 feet (in \cite{DO_396}, this distance is the threshold of a \emph{Loss of Well-Clear} event, or LoWC; and this distance is also prescribed in the standard ASTM F3442/F3442M-20 \cite{ASTM}). The analogous parameter in DO-365B has the value of 4,000 feet or 0.66 nmi (according to Table 2-24 of \cite{DO_365}, section ``HAZ'', row ``HMD''), and is part of the definition of \emph{DAA Well-Clear} or DWC. It may be possible to tweak their parameters to make them more comparable, but they have dozens of other parameters that are not directly translatable and, furthermore, ACAS sXu has large look-up tables. Thus, we kept most of their original parameters values intact and consider any comparison between them as just notional.   

%% file: scenario_definition.tex
\begin{figure}[ht!]
	\begin{center}
		\includegraphics[width=0.65\columnwidth]{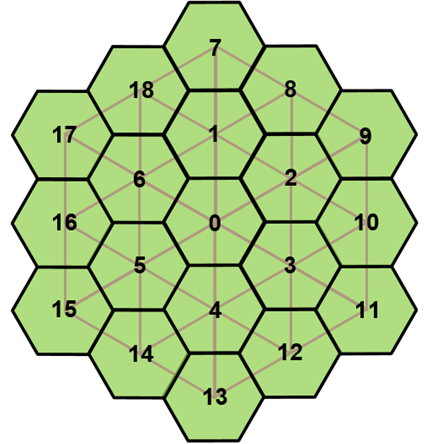}
		\caption{Cellular structure of the airspace, with cells of 2,000 m of radius, on the horizontal plane.}
		\label{fig:cellular_airspace}
	\end{center}
\end{figure}

The multi-vehicle scenarios consist of an airspace of approximately 20 km of diameter, with 4 aircraft and definitions similar to \cite{IROliveira2021}, 
but here, one of the cases allows 3-D maneuvering. In this section, we present a summary of these common scenario definitions. The airspace was initially designed
with a cellular structure, in order to facilitate coordination among the agents, as shown in  Fig.~\ref{fig:cellular_airspace}.

In this paper's analyses, the cells are not needed for coordination, however that structure is still needed to generate the large number of different traffic configurations, because the origin and destination points of an air vehicle's mission are placed in the outer cells (7-18). The traffic configurations are generated by a factorial algorithm with the following constraints:
\begin{enumerate}
	\item \label{itm:aircraft_id} An aircraft is uniquely identified by a numerical id between 0 and 3;
	\item An aircraft's mission is defined by an Origin and a Destination point, both being cell centroids;
	\item Origin points are all distinct among the aircraft, so as that they begin the scenario properly separated;
	\item \label{itm:o-d_distance} The Destination point must be distant from the Origin point by at least three cells in-between them;
\end{enumerate}
Constraint \#\ref{itm:aircraft_id} allows to define priority rules among the aircraft, and contributes to generate a higher number of distinct traffic configurations. Constraint \#\ref{itm:o-d_distance} is thought to exclude trajectories that are too short and, in a certain way, to promote convergence and crossing of the trajectories and obtain more conflicts. Implicitly, it forces that the origin and destination points be at the outer ring of the airspace. Then, by generating all possible combinations allowed by these constraints, one obtains 122,416 distinct traffic configurations. Still, it can be observed that some of these configurations are rotated versions of others, so, if the cell numbers did not matter, some of these configurations could be eliminated and we would remain with 46,660 traffic configurations (the full set does not contain 6 rotated replica subsets because constraint \#\ref{itm:o-d_distance} creates asymmetries). Anyway, we still chose to keep the full set because some conflict resolution algorithms use the reference to the north (and consequently, to the cell numbers) to decide tie-break situations, and even because some of them use look up tables or decision rules in which the absolute heading of the aircraft is an input. In some of our experiments, it was observed that ACAS sXu has slightly different resolution to traffic configurations that are rotated versions, besides occurrence of special events with DAIDALUS happening for just one scenario among the total 122,416, confirming that there is no strong rotational symmetry in the DAA implementations. 

An aircraft can go out of the cell area to execute a deviation maneuver, and we observed such occurrences with a small frequency, depending on the scenario specification. In reality, there might be borders with forbidden airspaces or obstacles, and those would be more significant safety threats. In any case, the aircraft tend to remain inside the airspace because they have to reach their destination through an efficient path. For the scenario specifications with only vertical maneuvers, the aircraft remain symbolically at 500 ft of altitude, while in the spec with vertical maneuvers, the aircraft have their origin and destination points at this altitude, and there will remain if no vertical maneuver happen. Vertical maneuvers are allowed only above this altitude, without any bound up, other than the requirement to return to the base altitude in the end of the flight mission. 

\subsection{Use of priorities}

Priorities can be used to determine a collective decision on how each aircraft will alter its trajectory in case of a traffic encounter of two or more aircraft \cite{ROMANIDEOLIVEIRA201792}. DAA attributes priorities to each aircraft in an encounter according to built-in, safety-driven rules. Although DO-365B \cite{DO_365} does not establish any decision rule based on priority, it requires a high priority flag indication for the remote pilot in case of an intruder is in emergency state. On the other hand, both DAIDALUS \cite{DaidalusPage} and DO-386 \cite{DO_386} define, respectively, priority rules for coordinated encounters, where priorities are defined according to observed aircraft equipage and the  aircraft identification code (the so-called ICAO address). The intention is to use objective rules to decrease the chances of aircraft performing maneuvers that worsen the conflict, as for example, turning onto each other. We distinguish these built-in priorities from the priorities explored in \cite{ROMANIDEOLIVEIRA201792}, by calling the latter as \emph{extrinsic} priorities, because they can be arbitrarily chosen among equal vehicles. And it is necessary bearing in mind that the DAA built-in priorities are used only for the divergent maneuvers, not for the return-to-mission phase. 

Considering the case where aircraft are equally equipped, DAA maneuver tiebreaking is performed by aircraft id, and is fair only if the probability distribution of aircraft ids per encounter follow a symmetric curve, which would hardly be the case in practice. Apart from tiebreaking, there is a general understanding in the DAA regulations that priorities must be observed when encounters occur with public service and medevac missions, vehicles in various sorts of emergency, etc. However, the afore mentioned standards do not need to define specific rules for encounters with such intruders, because they assume that such intruders will be treated as non-cooperative targets, either by proactive identification by the remote pilot, or by interaction with Air Traffic Control, or, as the last means, by the repeated refusal of cooperation by the intruder. 

It was observed in our previous works \cite{ROMANIDEOLIVEIRA201792,IROliveira2021} that, for encounters involving multiple (more than two) equally-righted air vehicles in dense airspaces, the use of extrinsic priorities improves the efficiency of the collective traffic system, because it decreases the total number of deviations and gives more predictability to the scenarios. In the case of DAA, they can be used to make DAA to suppress any alert associated to a cooperative intruder with lower priority than the ownship's. While this greatly improves efficiency of the scenarios, it also may increase the occurrence of losses of separation, and this trade-off has to be taken into account. Such trade-off also includes the occurrence of timeout events, already explored in our previous works. 

A timeout is the forced termination of a long chain of maneuvers without all the aircraft having accomplished their missions. In practice, this would be most likely associated with a fuel or energy emergency, when the aircraft's energy supply becomes critically low. Thus, at least one of the aircraft has to start a procedure of emergency landing, which is an undesirable event, both from the efficiency and the safety perspectives. The long chain of maneuvers is caused by the inability of the DAA logic to plan a coordinated return to mission for all the aircraft involved in an encounter, thus each aircraft decides independently on how to resume its intended path, and this lack of coordination causes maneuver repetitions, extensions, alternating cycles, and other bizarre sequences of maneuvers, that could also be called \emph{livelocks}. 

In some of our scenario specifications, we used the aircraft id as the sole criterion for extrinsic priorities. This choice is technically correct, however, in practice, it is not  an equitable policy, because the id remains unchanged for the lifetime of the vehicle, and that either blesses or condemns the vehicle with a persistent tendency to being promoted or being penalized. Thus, a randomized priority assignment would be needed, but that would have to be cheat-proof and fault-proof \cite{IROliveira2022}, which would bring some extra cost to the system.  

\subsection{Vertical maneuvering}

Most of the scenarios that we studied have only horizontal maneuvers, with all aircraft flying at the same altitude of 500 ft. This might happen in practice near airports, where the higher altitudes are reserved for manned aviation, and lower altitudes are not allowed due to ground obstacles or noise regulations. In any case, 2-D aircraft dynamics and mission control is much simpler and greatly helps in grasping the critical aspects of distributed Conflict Detection and Resolution (CD\&R). Later in this research, we developed an initial 3-D simulation model in which, if DAA presents avoidance bands in the vertical dimension, only the up sense is taken, a decision depending on a priority order for choosing which of the aircraft will maneuver or not. This was a simplification to treat the indeterminacy left by DO-365B and DAIDALUS. DO-365B associates coordinated encounters to certain classes of equipage, among which the Class 1 does not allow such coordination. 

In DAIDALUS, there is no explicit coordination among the aircraft and, because of that, the ownship ``pilot" can choose either direction and end up choosing the same direction of the intruder ``pilot". Thus, its resolution logic will be complete only with additional assumptions. It is easy to see that, in the vertical plane, if both aircraft maneuver up for conflict avoidance, there will be no avoidance. In the horizontal plane, where the aircraft dynamics has rotational symmetry, it is desirable that aircraft turn to the same relative direction, however sensor uncertainty and multi-aircraft encounters remain problematic without coordination. In our implementation of 3-D maneuvering, if DAIDALUS is prescribing an upwards vertical speed, that becomes a control input simultaneously to any horizontal deviation input, thus the resulting maneuver can happen in both planes at the same time. Once DAIDALUS signals that the aircraft is clear of conflict, the mission control logic will try to return to the destination altitude, provided that it does not create a new conflict immediately. With regards to selection of upward or downward resolution maneuver, the use of id-based priorities allowed our implementation to disregard downward advisories with acceptable results and solve the vertical indeterminacy in a simplified way.  For an in-depth study on the interoperability between horizontal and vertical advisories, one can refer to \cite{Londner2015}.

%% file: analysis.tex
\subsection{Performance Indicators}\label{subsection:performance_indicators_definition}
We use three categories of indicators to analyze the performance of a scenario definition:

\textbf{Inefficiency Rate}: it is the difference in fuel spent between a traffic scenario in which the aircraft comply to DAA resolution advisories (RA), and the analogous scenario in which each vehicle follows its optimal path, without observation of traffic separation rules. In case of vertical deviations, a higher fuel rate is required for climb maneuvers, a lower fuel rate in descent maneuver, which increase the net total for the mission. The resulting value of this indicator is the average value over all traffic configurations in a scenario set. 

\textbf{Loss of Separation (LoS) Rate}: despite there being several ways of defining traffic separation, we examine just the simplest one, which is checking whether or not the vehicles are separated by at least a fixed \emph{minimum distance}. In order to include both DAIDALUS and ACAS sXu in the same tables, we use one separation distance from each one, respectively: 4,000 ft, which is the Horizontal Miss Distance, or HMD, used in DO-365B \cite{DO_365} to define the so-called \emph{Hazard Alert Zone} (HAZ), associated to the DAA Well-Clear (DWC) concept of separation; and 2,000 ft, used in DO-396 \cite{DO_396}, that defines the Loss of Well-Clear (LoWC) event in relation to large UAVs or manned aircraft. These indicators will denote the rate of scenarios, in a scenario set, where the distance between any aircraft pair fell below the afore mentioned threshold values. 

\textbf{Timeout Rate}: this indicates, in a scenario set, the rate of scenarios where any aircraft exceeded a maximum time without reaching its destination point. As pointed out in section \ref{section:scenario_definitions}, this phenomenon occurs because DAA Resolution Advisories (RAs) cause long chains of maneuvers that extend beyond the energy/fuel allowance of the vehicle, due to shortcomings in coordination. We use a time threshold of 1,000 seconds but, in practice, the timeout would be determined by the energy/fuel capacity of the vehicle.

\textbf{Scenario Computing Time}: the time needed to simulate a single scenario instance, in a single core of an Intel Xeon CPU, discounted the fact that multiple scenario instances can be run in parallel in a multi-core CPU. In our simulated scenarios, the DAA algorithm is called at least each 2 seconds, for each aircraft, but when the aircraft is in avoidance mode, that can happen more often. In the case of ACAS sXu, the requirement of receiving various messages to update a single track contribute to result in multiple calls per simulated second.

\subsection{Scenario Specifications and Labels}
In this study, a scenario specification is defined by features such as: the DAA algorithm used, the dimensionality (2-D or 3-D), if it uses extrinsic priorities or not, the target separation parameter, and possibly other features. The scenario labels used in this section encode these attributes:
\begin{itemize}
\item \texttt{dai\_ip\_2d\_4k}: DAIDALUS without extrinsic priorities, 2-D maneuvering and regular 4 kft Horizontal Miss Distance (HMD);
\item \texttt{dai\_ep\_2d\_4k}: similar to the above, with extrinsic priorities;
\item \texttt{sxu\_ip\_2d\_2k}: ACAS sXu with intrinsic priorities only, 2-D maneuvering and regular 2 kft LoWC threshold;
\item \texttt{sxu\_ep\_2d\_2k}: similar to the above, with extrinsic priorities;
\item \texttt{dai\_ep\_3d\_4k}: similar to \texttt{dai\_ep\_2d\_4k}, with 3-D maneuvering;
\item \texttt{dai\_ep\_2d\_2k}: similar to \texttt{dai\_ep\_2d\_4k}, with HMD reduced to 2 kft.
\end{itemize}

And there are other features and labels that will be mentioned below as needed.

\subsection{Performance Analysis}
The analysis of selected scenario specifications is summarized in table~\ref{table:performance_analysis}. The first notorious observation in this table is the effect of extrinsic priorities to decrease inefficiency. With regards to safety indicators, their effect is mixed, and we have to observe each case separately. In the case of DAIDALUS, priorities decreased the 2 kft LoS rate and, most drastically, the timeout rate, while increased the 4 kft LoS rate. In the case of ACAS sXu, priorities increased both LoS rate indicators, but decreased the timeout rate drastically. Based on our rule of thumb assessment, it can be said that DAIDALUS works better with extrinsic priorities, while ACAS sXu works better without them. We conjecture that the following reasons explain this fact: i) that DAIDALUS has more built-in symmetries than ACAS sXu; ii) that ACAS sXu has already built-in priority rules for multi-aircraft encounters, and extrinsic priorities may contradict with them.

\begin{table}[h]
    \caption{Summary of closed-loop performance indicators per scenario.}
    \label{table:performance_analysis}
    \begin{center}
    \begin{tabularx}{\columnwidth}{|p{0.20\columnwidth}|p{0.11\columnwidth}|p{0.09\columnwidth}|p{0.09\columnwidth}|p{0.1\columnwidth}|p{0.105\columnwidth}|}
        \hline
    Scenario spec. & Ineffici-ency rate & LoS rate 4~kft & LoS rate 2~kft & Timeout rate & Scenario comp. time (s)\\
    \hline
    \texttt{dai\_ip\_2d\_4k} & 9.71\% & 1.4E-2 & 6.5E-5 & 1.6E-2 & 6.5E-2\\
    \texttt{dai\_ep\_2d\_4k} & 4.83\% & 2.4E-2 & 4.1E-5 & 0 & 5.1E-2\\
    \texttt{sxu\_ip\_2d\_2k} & 20.7\% & 8.7E-1 & 4.1E-2 & 1.6E-3 & 7.8E+1\\
    \texttt{sxu\_ep\_2d\_2k} & 10.9\% & 9.0E-1 & 2.0E-1 & 1.8E-5 & 7.8E+1\\
    \texttt{dai\_ep\_3d\_4k} & 4.38\% & 8.9E-6 & 0 & 0 & 7.4E-2\\
    \texttt{dai\_ep\_2d\_2k} & 1.3\% & 9.0E-1 & 1.1E-1 & 0 & 3.9E-2\\
    \hline
    \end{tabularx}
    \end{center}
\end{table}

According to a line of reasoning, it would be expected, that, in the more efficient scenarios, the aircraft fly closer to each other and, therefore, there should be a higher probability of losing separation. But this is not the only principle at play, because, if the aircraft perform deviations with the least extra distance, while keeping separation, they stay less in the air and decrease the total number of conflicts. This becomes more understandable when we compare \texttt{dai\_ep\_2d\_4k} with \texttt{dai\_ep\_3d\_4k}, where the latter achieved a small advantage in efficiency, but a huge one in safety. \texttt{dai\_ep\_3d\_4k} is capable of shortening the total distances, but has a residual cost associated to vertical maneuvers, where the climb maneuvers spend fuel at higher rates. 

It cannot escape from observation that DAIDALUS performed much better than ACAS sXu in almost all indicators. In our opinion, it would be reasonable to expect that ACAS sXu would not excel in the LoS rates, especially that of 4 kft, because its first protection criterion is 2 kft, as a built-in feature. However, with smaller protection volumes, the deviations should be smaller and, by this reasoning, its expected inefficiency would be lower than that of DAIDALUS. But our results show otherwise when we compare the cases of DAIDALUS with those of ACAS sXu. The only case in which ACAS sXu obtained an advantage was for the LoS rates comparison between \texttt{sxu\_ip\_2d\_2k} and \texttt{dai\_ep\_2d\_2k}, which have the same separation target. In any case, the Los rate obtained for ACAS sXu meets the performance requirement of ASTM F3442 \cite{ASTM}, which uses the definition of LoWC Ratio (LR), which is the ratio between the LoS rate of 2 kft shown in table~\ref{table:performance_analysis}, with DAA active, and corresponding LoS rate with DAA inactive, the latter being 0.785 according  to our simulations. Thus, the resulting LR scores for ACAS sXu here are 0.052 and 0.253, respectively for the two ACAS sXu specs, which are well below the value of 0.4 from \cite{ASTM}, and consistent with the performance analysis of \cite{DO_396}. We conjecture that these scores would be lower in a future 3-D scenario spec of a ACAS sXu, by following the same improvement obtained with DAIDALUS. 

\subsection{Possible approximations to the closed-loop behavior}
\begin{figure}[ht!]
	\begin{center}
		\includegraphics[width=1.0\columnwidth]{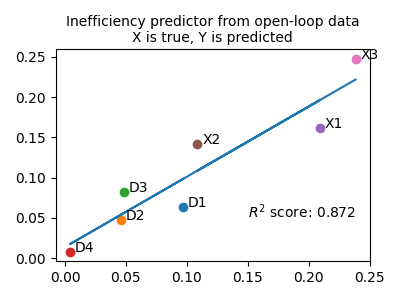}
		\caption{Prediction of closed-loop inefficiency based on open-loop  $\overline{M/D}$ and $\overline{\alpha}$ measures.}
		\label{fig:inefficiency_regression}
	\end{center}
\end{figure}

Trying to alleviate the heavy computational load to simulate large numbers of different traffic configurations, in this multi-aircraft, closed-loop setup, we considered some approximated solutions, such as the use of Deep Neural Networks to emulate the ACAS Xu/sXu behavior, in the lines followed by \cite{Julian2018,Bak2022}. However, the existing solutions that we found available were developed for just one intruder aircraft, so they were not suitable for our study. Another possibility would be the exploitation of symmetry transformations \cite{Sibai2020}, however the history-dependent nature of the ACAS Xu/sXu algorithms, associated to the present closed-loop setup, make this possibility unpractical. Thus, we started exploring simpler ways of deducing closed-loop behavior without having to perform the full simulation of a scenario. So far, we tried to analyze correlations between measures of open-loop maneuvers and the closed loop performance. The features that we explored are:
\begin{itemize}
    \item \textbf{Distance flown until the end of the first deviation maneuver} ($\overline{M/D}$): we consider the total distance flown until a ``Clear-of-Conflict'' (CoC) event happens, that is, after one or more divergent maneuvers start in a scenario instance, we stop the scenario when the first divergent maneuver of any aircraft finishes and that individual aircraft is clear of conflict, the moment from which some decision must be made on how to continue the mission. We count the total number of maneuvers started, and divide it by the sum of the flown distances, across all scenario instances in an execution set associated to a scenario spec.
    \item \textbf{Average angle deviation maneuver} ($\overline{\alpha}$): using the same stopping rule of above, we account the angle difference between the heading angles of the aircraft at the beginning of the divergent maneuver and at the stopping moment. 
\end{itemize}
For each of these measures, we ran the 122,416 traffic configuration instances with the open-loop stopping rule. Here, all the scenario specifications are 2-dimensional, and we use abbreviated lables to achieve a better display in the graph legends. Namely, the scenario specifications in this subsection are defined as:
\begin{itemize}
    \item \texttt{D1}: DAIDALUS without extrinsic priorities and with deterministic sensor data;
    \item \texttt{D2}: DAIDALUS with extrinsic priorities and deterministic sensor data;
    \item \texttt{D3}: DAIDALUS with extrinsic priorities and Sensor Uncertainty Mitigation (SUM). This is a design feature \cite{Narkawicz2018} to mitigate uncertainty in sensor data, as for example, to determine the position of an intruder aircraft;
    \item \texttt{D4}: DAIDALUS with its Horizontal Miss Distance (HMD) set to 200 ft (the standard is 4,000 ft) and uncertain sensor data;
    \item \texttt{X1}: ACAS sXu without extrinsic priorities;
    \item \texttt{X2}: ACAS sXu with extrinsic priorities;
    \item \texttt{X3}: ACAS sXu with scenario downscaled to speed of 43 knots and cell radius of 1 km.
\end{itemize}
We used the values of $\overline{M/D}$ and $\overline{\alpha}$ obtained for each of the specs above as inputs to a linear regressor of inefficiency, as defined in section~\ref{subsection:performance_indicators_definition}, and generated a plot with the pairs of (true, predicted) values from this regression, as shown in fig.~\ref{fig:inefficiency_regression}. The trend line in the figure, which depicts the regressor, seems to represent a strong correlation, which is confirmed by the value of $R^2$. When considering each of the regression inputs separately, we obtain $R^2=0.73$ for $\overline{M/D}$ and $R^2=0.77$ for $\overline{\alpha}$, which show that they contribute with approximately equal predicting power. 

We performed a similar analysis for the LoS indicators, but we found very little correlation, as $R^2=0.11$ for the 2~kft LoS indicator. Nevertheless, it can be concluded that these open-loop measurements are a good proxy for the closed-loop inefficiency, with the advantage that the predictor discounts the bias that may have been introduced by the closed-loop mission management system, which is not part of the DAA specification. A rough estimate for the computing time saved is of 68\% in the inefficiency case.

%% file: conclusion.tex
In this paper, we presented an extension of our work on the analysis of closed-loop air traffic management algorithms, aimed at fully autonomous applications. This extension focused solely on DAA algorithms, using two implementations fulfilling industry standards, namely DAIDALUS and ACAS sXu. We analyzed the effect of using extrinsic (or arbitrarily decided) priorities as a means of coordination, and found that they cause a drastic decrease in inefficiency and timeout rates, for all algorithms, with a positive effect also for the rate of Loss of Separation (LoS) in the case of DAIDALUS with the threshold of 2,000 ft prescribed by the standard ASTM F3442 \cite{ASTM}, while in the case of ACAS sXu it had a negative effect on LoS rate. Our results show that ACAS sXu has higher inefficiency and LoS rates than DAIDALUS in physically similar scenarios, apart from considerations on noise and errors in sensor data. This higher inefficiency of ACAS sXu is confirmed by prediction based on open-loop scenario data, in which the aircraft perform maneuvers based solely on DAA advisories. The use of open-loop indicators for predicting inefficiency showed to be promising, with a high degree of correlation registered for a small number of scenario specifications, while the prediction of LoS rates based on open-loop simulation turned out to have little correlation. The use of only open-loop data allows to save 68\% of simulation time. 

A final confirmation remains to be done in this phase of the research, which is to integrate the vertical maneuvers of ACAS sXu with our closed-loop mission control logic, in order to obtain the performance of the complete ACAS sXu logic in 3-D. Because this will certainly increase the already massive computational effort for the many traffic configurations, we also need to find an unbiased sampling rule to decrease the number of scenario instances, or perhaps develop a Deep Neural Network model of ACAS Xu/sXu for 3 or more intruders. Besides that, the difficulties that we found in achieving an efficient decentralized strategy for traffic management subject to DAA advisories raised the need of finding more powerful methods to solve this problem, and the method that seems more promising, based on a preliminary literature search, is Multi-Agent Reinforcement Learning (MARL), which we intend to explore in the next steps. 

%% file: closed_loop_DAA_2023.bbl
\begin{thebibliography}{10}
\providecommand{\url}[1]{#1}
\csname url@samestyle\endcsname
\providecommand{\newblock}{\relax}
\providecommand{\bibinfo}[2]{#2}
\providecommand{\BIBentrySTDinterwordspacing}{\spaceskip=0pt\relax}
\providecommand{\BIBentryALTinterwordstretchfactor}{4}
\providecommand{\BIBentryALTinterwordspacing}{\spaceskip=\fontdimen2\font plus
\BIBentryALTinterwordstretchfactor\fontdimen3\font minus
  \fontdimen4\font\relax}
\providecommand{\BIBforeignlanguage}[2]{{%
\expandafter\ifx\csname l@#1\endcsname\relax
\typeout{** WARNING: IEEEtran.bst: No hyphenation pattern has been}%
\typeout{** loaded for the language `#1'. Using the pattern for}%
\typeout{** the default language instead.}%
\else
\language=\csname l@#1\endcsname
\fi
#2}}
\providecommand{\BIBdecl}{\relax}
\BIBdecl

\bibitem{DO_365}
``{Minimum Operational Performance Standards (MOPS) for Detect and Avoid (DAA)
  Systems},'' Radio Technical Commission for Aeronautics (RTCA), SC-147,
  \unskip\space DO-365B, Mar. 2018.

\bibitem{DO_386}
``{Minimum Operational Performance Standards for Airborne Collision Avoidance
  System Xu (ACAS Xu)},'' Radio Technical Commission for Aeronautics (RTCA),
  SC-147, \unskip\space DO-386, Dec. 2020.

\bibitem{DO_396}
``{Minimum Operational Performance Standards for Airborne Collision Avoidance
  System sXu (ACAS sXu)},'' Radio Technical Commission for Aeronautics (RTCA),
  SC-147, \unskip\space DO-396, Dec. 2022.

\bibitem{IROliveira2021}
I.~Romani~de Oliveira, E.~Pinto~Neto, T.~Matsumoto, H.~Yu, E.~Bartolom\'e,
  G.~Frontera, and A.~Mayne, ``{Comparing the Performance of Traffic
  Coordination Methods for Advanced Aerial Mobility},'' in \emph{2021 IEEE/AIAA
  40th Digital Avionics Systems Conference}.\hskip 1em plus 0.5em minus
  0.4em\relax IEEE, Oct. 2021.

\bibitem{DaidalusPage}
``{DAIDALUS},'' \url{https://shemesh.larc.nasa.gov/fm/DAIDALUS/}, last accessed
  29-March-2023.

\bibitem{Manfredi2016}
G.~Manfredi and Y.~Jestin, ``An introduction to acas xu and the challenges
  ahead,'' in \emph{2016 IEEE/AIAA 35th Digital Avionics Systems Conference
  (DASC)}, 2016, pp. 1--9.

\bibitem{Owen2019}
M.~P. Owen, A.~Panken, R.~Moss, L.~Alvarez, and C.~Leeper, ``Acas xu:
  Integrated collision avoidance and detect and avoid capability for uas,'' in
  \emph{2019 IEEE/AIAA 38th Digital Avionics Systems Conference (DASC)}, 2019,
  pp. 1--10.

\bibitem{Davies18}
J.~T. Davies and M.~G. Wu, ``{Comparative Analysis of ACAS-Xu and DAIDALUS
  Detect-and-Avoid Systems},'' NASA, Technical Memorandum TM–2018–219773.

\bibitem{ASTM}
``{Standard Specification for Detect and Avoid System Performance
  Requirements},'' ASTM International, \unskip\space ASTM F3442/F3442M-20,
  2020.

\bibitem{ROMANIDEOLIVEIRA201792}
\BIBentryALTinterwordspacing
Ítalo {Romani de Oliveira}, ``Analyzing the performance of distributed
  conflict resolution among autonomous vehicles,'' \emph{Transportation
  Research Part B: Methodological}, vol.~96, pp. 92--112, 2017. [Online].
  Available:
  \url{https://www.sciencedirect.com/science/article/pii/S0191261516304854}
\BIBentrySTDinterwordspacing

\bibitem{IROliveira2022}
\BIBentryALTinterwordspacing
I.~Romani~de Oliveira, T.~Matsumoto, and E.~Pinto~Neto, ``{Blockchain-based Air
  Traffic Management for Advanced Air Mobility},'' in \emph{SITRAER 2022 Air
  Transportation Symposium}. [Online]. Available:
  \url{https://arxiv.org/abs/2208.09312}
\BIBentrySTDinterwordspacing

\bibitem{Londner2015}
E.~H. Londner, ``Interoperability of horizontal and vertical resolution
  advisories,'' in \emph{{Eleventh USA/Europe Air Traffic Management Research
  and Development Seminar (ATM2015)}}, 2015.

\bibitem{Julian2018}
\BIBentryALTinterwordspacing
K.~D. Julian, M.~J. Kochenderfer, and M.~P. Owen, ``Deep neural network
  compression for aircraft collision avoidance systems,'' \emph{Journal of
  Guidance, Control, and Dynamics}, vol.~42, no.~3, pp. 598--608, 2019.
  [Online]. Available: \url{https://doi.org/10.2514/1.G003724}
\BIBentrySTDinterwordspacing

\bibitem{Bak2022}
S.~Bak and H.-D. Tran, ``Neural network compression of acas xu early prototype
  is unsafe: Closed-loop verification through quantized state
  backreachability,'' in \emph{NASA Formal Methods}, J.~V. Deshmukh,
  K.~Havelund, and I.~Perez, Eds.\hskip 1em plus 0.5em minus 0.4em\relax Cham:
  Springer International Publishing, 2022, pp. 280--298.

\bibitem{Sibai2020}
H.~Sibai, N.~Mokhlesi, C.~Fan, and S.~Mitra, ``Multi-agent safety verification
  using symmetry transformations,'' in \emph{Tools and Algorithms for the
  Construction and Analysis of Systems}, A.~Biere and D.~Parker, Eds.\hskip 1em
  plus 0.5em minus 0.4em\relax Cham: Springer International Publishing, 2020,
  pp. 173--190.

\bibitem{Narkawicz2018}
A.~Narkawicz, C.~Muñoz, and A.~Dutle, ``Sensor uncertainty mitigation and
  dynamic well clear volumes in daidalus,'' in \emph{2018 IEEE/AIAA 37th
  Digital Avionics Systems Conference (DASC)}, 2018, pp. 1--8.

\end{thebibliography}
